\documentclass[prc,aps]{revtex4}
\draft
\usepackage{graphicx}
\usepackage{color}
\usepackage{mathtools}
\begin{document}

\title{Revised analysis of the strengths of the contact potential at N$^4$LO through nuclear and neutron matter }  
\author{            
Francesca Sammarruca\footnote{To Rup, in Loving Memory}\footnote{Corresponding author. Email: fsammarr@uidaho.edu} and Tomiwa Ajagbonna}                                                           
\affiliation{ Physics Department, University of Idaho, Moscow, ID 83844-0903, U.S.A. 
}
\date{\today} 
\begin{abstract}
\noindent
 We examine the contact three-nucleon force at N$^4$LO expressed as a density dependent potential. In Ref.~\cite{WGS22}, the necessary couplings (13, including two that appear in the leading three-nucleon force), were extracted from $nd$ scattering observables. The contact strengths obtained through the three-nucleon continuum, without fits to the triton, seem incompatible with the energy of nuclear and neutron matter. 
We take the opportunity to revisit the role of 
 the $c_D, c_E$ couplings of the leading three-nucleon force in nuclear matter and nuclei. We close with a discussion about recent fittings of three-nucleon scattering observables using a spectroscopic basis for the subleading contact three-nucleon force. Using the errors given for the couplings in the spectroscopic basis, we derive uncertainty bands for the strengths in the operator basis and their contributions to the leading and subleading three-nucleon contact interactions.
\end{abstract}
\maketitle 

\section{Introduction}
The role of the contact three-nucleon force (3NF) at N$^4$LO has been discussed extensively over the past 20 years, especially as a potential solution to a long-standing problem with the three-nucleon (3N) analyzing power -- the ``$A_y$ puzzle."~\cite{EMW02}. The interest in this 3NF contribution is motivated by its rich spin and isospin strucure, which brings in 13 additional couplings and thus promises much needed flexibility. Actually, two of these 13 strengths are related, making the total number of short-range couplings equal to 11+2, including the $c_D$ and $c_E$ LECs of the leading short-range 3NF (the 3NF at N$^3$LO is parameter free).

\begin{figure*}[!t] 
\centering
\hspace*{-0.5cm}
\includegraphics[width=8.0cm]{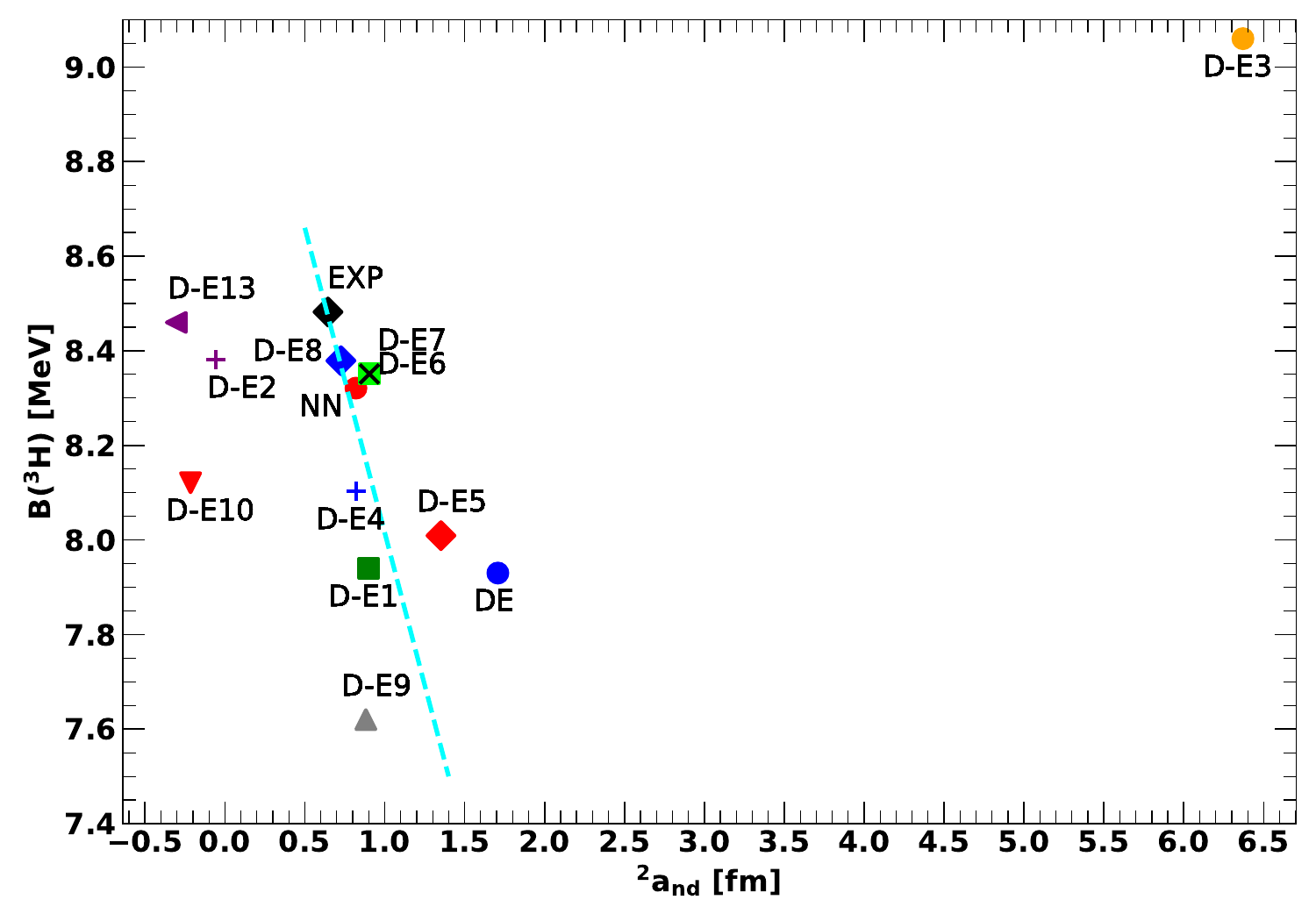}\hspace{0.01in}  
\vspace*{-0.1cm}
 \caption{Figure reproduced from Fig.~13 of Ref.~\cite{WGS22}. The black diamond marks the experimental values of the triton binding energy (8.4820(1) MeV) and the doublet scattering length (0.645 $\pm$ 0.003 fm). Starting with only the 2NF from the SMS N$^4$LO$^+$ potential~\cite{RKE18}, inclusion of the leading short-range 3NF with $c_D$ and $c_E$ as obtained from 3N scattering takes the result down to the blue circle. The other symbols show the impact of adding, consecutively, the $E_i$ contributions.
}
\label{WitFig}
\end{figure*}

In Ref.~\cite{WGS22}, these 13 parameters were obtained
by least squares fitting of theoretical predictions to cross section and analyzing powers data at
three energies of the projectile nucleon. The fitted strength
parameters were found to improve the description of elastic Nd scattering observables over a a wide range of incident energies, from 10 to 250 MeV. However, the improved description of the 3N continuum does not yield, at the same time, good predictions for the triton binding energy and the doublet $nd$ scattering length. This can be seen from Fig.~\ref{WitFig}, which we reproduced from Fig.~13 of Ref.~\cite{WGS22}. The figure displays a large spread on a $B(^3H)$--$^2a_{nd}$ plane when individual contact terms are included one by one, starting from the two-nucleon force (2NF) only, for which the SMS N$^4$LO$^+$ potential~\cite{RKE18} is employed. 

Note that the scenario is quite different in the two-nucleon (2N) sector, where high precision description of the 2N continuum yields, simultaneously, a good description of the 2N bound state. 
Part of the reason for the unsatisfactory description of the triton could be the absence of the 3NF at N$^3$LO~\cite{WGS22}.
Nevertheless, the authors conclude that the N$^4$LO contacts are promising and likely to be highly consequential for the spectra of nuclei.

In this work, we examine the contact 3NF at N$^4$LO through symmetric nuclear matter (SNM) and neutron matter (NM). Traditionally, SNM has been considered to be the ``test bench" for nuclear
many-body theories. As nuclear matter is the result of an extrapolation from finite nuclei, the saturation properties of SNM, that is, the minimum of the equation of state at the appropriate equilibrium density, should be naturally related to the energy and density distributions of nucleons in nuclei. In recent years, the connection between the saturating behavior of infinite nuclear matter and the
description of medium-mass nuclei has been investigated based on state-of-the-art chiral two- and three-nucleon forces, see Ref.~\cite{SM20} and references therein. Typically, the spot light of these discussions falls on the role of the $c_D$ and $c_E$ couplings appearing in the leading short-range 3NF, the large differences among their values when fitted through different systems/observables, and the degree of softness needed in the 2NF for a good description of intermediate-mass nuclei.

Here, we address the question of how the new couplings at N$^4$LO impact SNM and NM. At this time, any such calculation is unavoidably incomplete, as we omit the 3NF at N$^3$LO, which is presently not on stable grounds, and of course the complete 3NF at  N$^4$LO is missing. Even so, we think it's important to pursue an investigation of these contributions from the nuclear matter perspective, to identify 
 systematics, if any, between attraction/repulsion of the various contact terms in infinite matter as compared to the 3N continuum, as well as potential patterns with regard to specific sensitivities. 
        
\section{Symmetric nuclear matter} 
\label{SNM} 

For the purpose of setting a scale, we start with the contact 3NF at leading order (N$^2$LO) in SNM, with respect to which the contact contribution at N$^4$LO is expected to be suppressed.

\subsection{Contact interaction at leading order}

To represent the 3NF, we use the density dependent potentials derived by Kaiser {\it et al.}. At leading order, the contribution from the contact term to the in-medium nucleon-nucleon (NN) interaction is (cfr. Eq.~(25) in Ref.~\cite{HKW10}):
\begin{equation}
V_{cts} = -\frac{c_E k_F^3}{f_{\pi}^4 \Lambda \pi^2} \; .
\label{vn2lo}
\end{equation}
First, we take the (dimensionless) constant $c_E$ equal to 0.13 from Ref.~\cite{DHS19}, where sets of values were extracted using the N$^2$LO450 and the N$^3$LO450 potentials from Ref.~\cite{EMN17} (among other interactions). Fitting the binding energy of the triton, the authors determined a trajectory in the $c_D, c_E$ plane, from which they selected a few best pairs using nuclear matter saturation as a constraint.

\begin{figure*}[!t] 
\centering
\hspace*{-0.5cm}
\includegraphics[width=5.5cm]{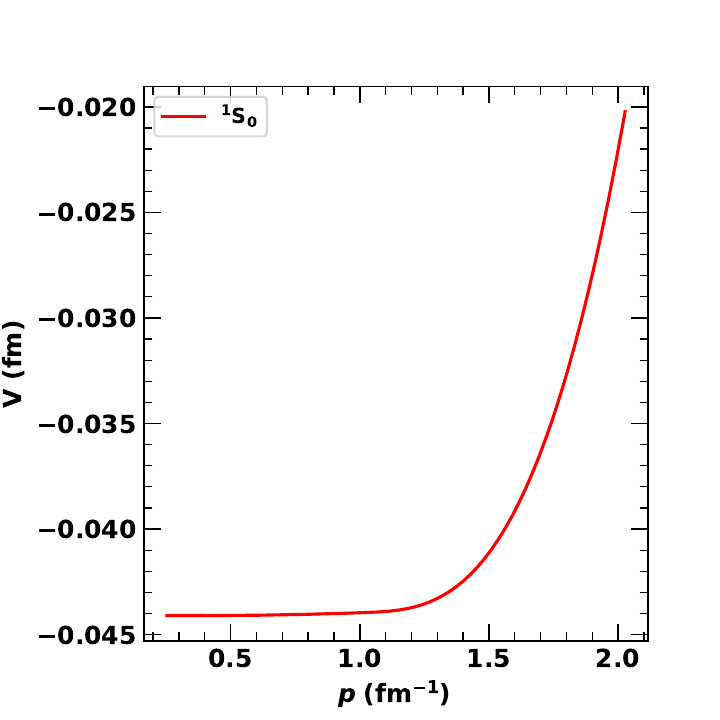}\hspace{0.01in} 
\includegraphics[width=5.5cm]{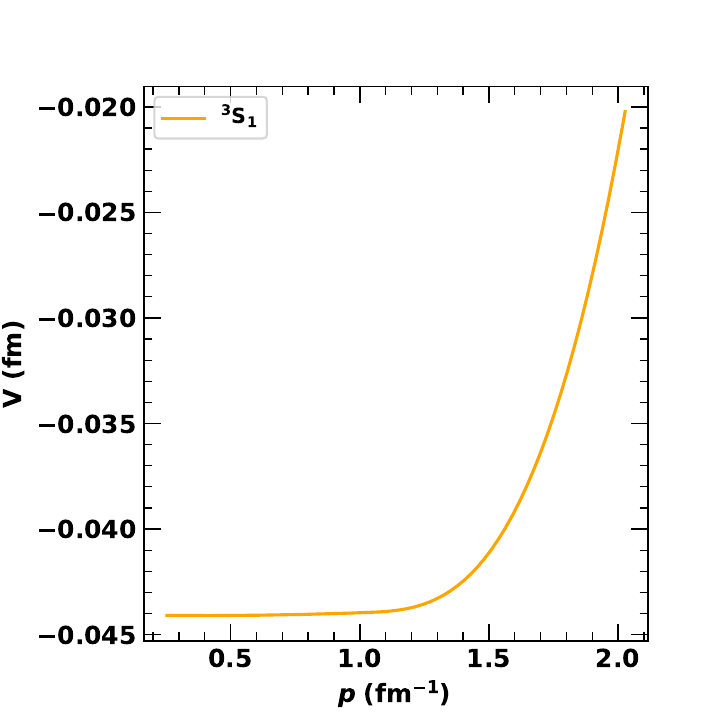}\hspace{0.01in} 
\includegraphics[width=5.5cm]{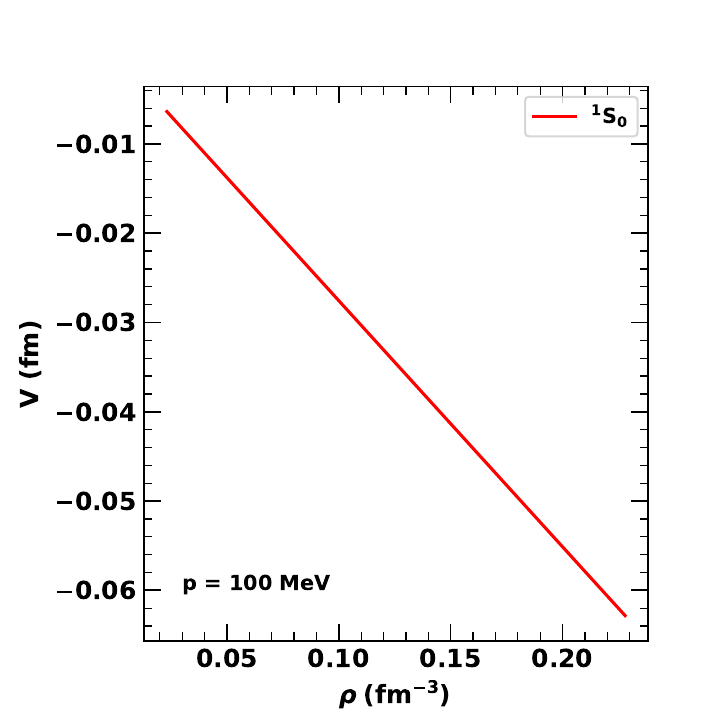}\hspace{0.01in} 
\includegraphics[width=5.5cm]{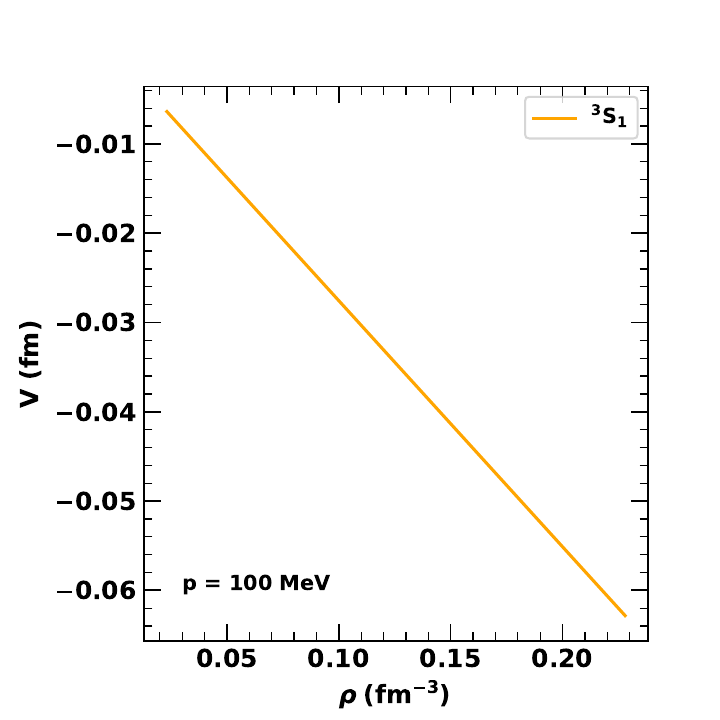}\hspace{0.01in} 
\vspace*{0.4cm}
 \caption{Matrix elements of Eq.~(\ref{vn2lo}) in $^1S_0$ and $^3S_1$ at $\rho$=0.16 $fm^{-3}$, as a function of momentum (first two frames from the left), and as a function of density at fixed momentum (next two frames).
}
\label{v6pw}
\end{figure*}

The top row of Fig.~\ref{v6pw} shows the matrix elements of Eq.~(\ref{vn2lo}) for $^1S_0$ and $^3S_1$ at normal density, $\rho$=$\rho_0$=0.16 $fm^{-3}$, as a function of momentum, while the bottom row displays the same states as a function of density at fixed momentum. The linear dependence on density is expected from Eq.~(\ref{vn2lo}), while momentum dependence is absent from the equation. The momentum dependence seen in the figures on the top row is due to the impact of the regulator, becoming noticeable above approximately 300 MeV.

\begin{table*}
\caption{Energy per particle in Born approximation vs. density. The third column shows the same quantity evaluated from Eq.(A5) of Ref.~\cite{Kaiser21}. The fourth column contains values obtained with $c_E$ =-1.27.}
\label{EvA}
\begin{tabular*}{\textwidth}{@{\extracolsep{\fill}}|cccc|}
\hline
\hline 
 $\rho$ ($fm^{-3}$) & E/A (MeV) &  E/A (MeV) & E/A (MeV)  \\
\hline
0.0232  & -0.0151 & -0.0151 & 0.148 \\
0.0346 & -0.0337 & -0.0337 & 0.329 \\
0.0492 & -0.0684 & -0.0684 & 0.668 \\
0.0675 & -0.1287 & -0.1287 &1.2571 \\
0.0899 & -0.2279 & -0.2280 & 2.2270 \\
0.1167 & -0.3840 & -0.3842 & 3.7536 \\
0.1484 & -0.6206 &-0.6211 & 6.0776 \\
0.160   & -0.7211 & -0.7219 & 7.0524 \\
0.166   & -0.7780 &-0.7789 & 7.610 \\
0.185  &  -0.9674 & -0.9690 & 9.4651 \\
0.228  & -1.4617 &  -1.4657 & 14.319 \\ 
\hline
\hline
\end{tabular*}
\end{table*}

Next, we estimate the contribution to the energy per nucleon in nuclear matter from the in-medium potential given in Eq.~(\ref{vn2lo}). The second column in Table~\ref{EvA}  is the result of our nuclear matter binding energy calculation in Born approximation. The third column is obtained from Eq.(A5) of Ref.~\cite{Kaiser21}, obviously equivalent to our Born calculation. The fourth column is obtained if using $c_E$=-1.27, extracted from the 3N continuum~\cite{WGS22}.

The contributions to the energy per particle in SNM from the short-range $c_E$-dependent in-medium potential differ dramatically when using values of $c_E$ constrained through different systems. In particular, the $c_E$ coupling extracted from 3N scattering data yields leading conributions to the energy that are opposite in sign and about one order of magnitude larger than if using values constrained as in Ref.~\cite{DHS19}.

\subsection{Contact interaction at N$^4$LO}
 Here, we examine the contributions from the N$^4$LO contact interaction.

The subleading 3N-contact potential is given by~\cite{Kai20_nn}:
\begin{equation}
\label{vmed}
\begin{split}
\frac{V_{med}}{\rho} & = E_1 \big (\frac{6}{5} k_f^2 +2 p^2 -3q^2) + E_2 \big [(\vec{\tau}_1 \cdot \vec{\tau}_2 + 3)  \big (\frac{3}{5} k_f^2 + p^2 \big ) - \vec{\tau}_1 \cdot \vec{\tau}_2 q^2 \big ]   \\
& + E_3 \big [ ( \vec{\sigma}_1 \cdot \vec{\sigma }_2 + 3)  \big (\frac{3}{5} k_f^2 + p^2 \big)  -  \vec{\sigma}_1 \cdot \vec{\sigma }_2 q^2 \big ] \\
 & + E_4 \big [(\vec{\tau}_1 \cdot \vec{\tau}_2  \vec{\sigma}_1 \cdot \vec{\sigma }_2 +9)  \big (\frac{3}{5} k_f^2 + p^2 \big) - \vec{\tau}_1 \cdot \vec{\tau}_2  \vec{\sigma}_1 \cdot \vec{\sigma }_2q^2 \big ] \\
& +(E_5 + \vec{\tau}_1 \cdot \vec{\tau}_2 E_6)  \big [ \vec{\sigma}_1 \cdot \vec{\sigma }_2 (q^2 - p^2) -3 \vec{\sigma}_1 \cdot \vec{q} \vec{\sigma}_2 \cdot \vec{q} +\frac{3}{2} (\vec{\sigma}_1 \cdot \vec{p} \vec{\sigma}_2 \cdot \vec{p} + \vec{\sigma}_1 \cdot \vec{p'} \vec{\sigma}_2 \cdot \vec{p'}) \big ] \\
&  + (7E_7 - 9E_8)\frac{i}{4}( \vec{\sigma}_1 + \vec{\sigma }_2)\cdot (\vec{q} \times \vec{p}) + E_9 \big [  \vec{\sigma}_1 \cdot \vec{q} \vec{\sigma}_2 \cdot \vec{q} +\frac{q^2}{2} - p^2 - \frac{3}{5} k_f^2  - \frac{i}{2} ( \vec{\sigma}_1 + \vec{\sigma }_2)\cdot (\vec{q} \times \vec{p})  \\
& + E_{10} \big [\vec{\tau}_1 \cdot \vec{\tau}_2  \vec{\sigma}_1 \cdot \vec{q} \vec{\sigma}_2 \cdot \vec{q}     + \frac{3q^2}{2} - 3p^2 - \frac{9}{5} k_f^2  - \frac{3i}{2} ( \vec{\sigma}_1 + \vec{\sigma }_2)\cdot (\vec{q} \times \vec{p}) \big ] \\
& + E_{11} \big [ \vec{\sigma}_1 \cdot \vec{q} \vec{\sigma}_2 \cdot \vec{q}   +  \frac{q^2}{2} - p^2 - \frac{3}{5} k_f^2 + \frac{i}{2} ( \vec{\sigma}_1 + \vec{\sigma }_2)\cdot (\vec{q} \times \vec{p}) \big ]  \\
& + E_{12} \big [ \vec{\tau}_1 \cdot \vec{\tau}_2 \vec{\sigma}_1 \cdot \vec{q} \vec{\sigma}_2 \cdot \vec{q}   +  \frac{3q^2}{2} - 3p^2 - \frac{9}{5} k_f^2 + \frac{3i}{2} ( \vec{\sigma}_1 + \vec{\sigma }_2)\cdot (\vec{q} \times \vec{p}) \big ]  \\ 
& + \frac{E_{13}}{2} \big \{ \vec{\tau}_1 \cdot \vec{\tau}_2 \big [q^2 - 2p^2  - \frac{6}{5} k_f^2 - \vec{\sigma}_1 \cdot \vec{q} \vec{\sigma}_2 \cdot \vec{q} + i ( \vec{\sigma}_1 + \vec{\sigma }_2)\cdot (\vec{q} \times \vec{p}) \big ] -3 \vec{\sigma}_1 \cdot \vec{q} \vec{\sigma}_2 \cdot \vec{q} \} \; .
\end{split}
\end{equation}

From Eq.(\ref{vmed}), the contribution to the energy per particle at first order is~\cite{Kaiser21}:
\begin{equation}
\label{EoA_kaiser}
\frac{E}{A} = \frac{k_f^8}{10 \pi ^4} (2E_1+2E_2+2E_3+6E_4-E_9-3E_{10}-E_{11}-3E_{12}+E_{13}) \; .
\end{equation}

\begin{figure*}[!t] 
\centering
\hspace*{-0.5cm}
\includegraphics[width=6.0cm]{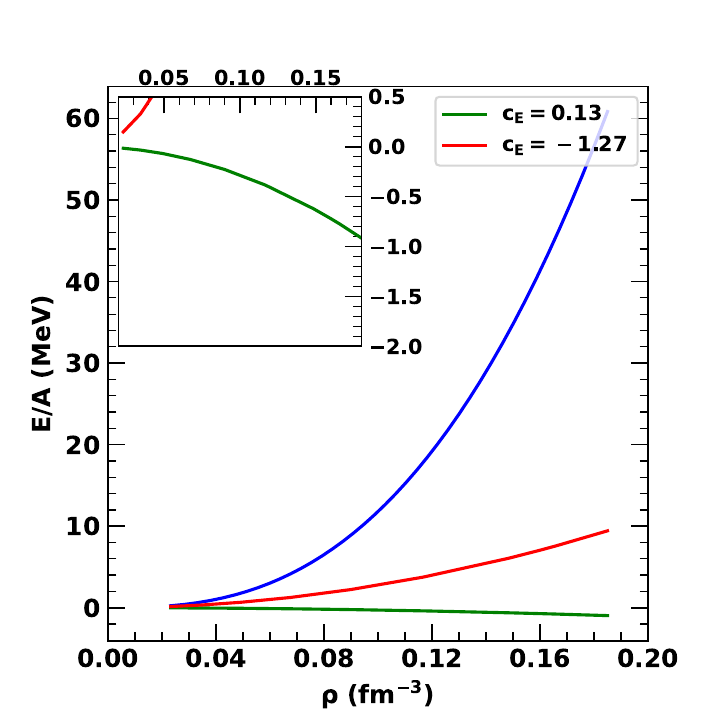}\hspace{0.01in}  
\vspace*{-0.1cm}
 \caption{Blue: Contribution to the energy per particle in Born approximation from the contact terms at N$^4$LO~\cite{Kaiser21}; Red: Contribution to the energy per particle in Born approximation from the contact term at N$^2$LO with $c_E$ = -1.27, from the 3N continuum; Green: As for the red curve, but with $c_E$ = 0.13~\cite{DHS19}. In the inset, the green curve is shown on an appropriate scale for better viewing.
}
\label{Kaiser1}
\end{figure*}   

\begin{table*}
\caption{A rough estimation of the contribution to the energy per particle in nuclear matter from the contact in-medium potential at N$^4$LO. }
\label{supr}
\centering
\begin{tabular}{|l|r|}
\hline
\hline
$\rho$ ($fm^{-3}$)  &  E/A (MeV) \\
\hline
0.0232  &  -0.000589 \\
  0.0346 &  -0.00172 \\
  0.0492 &  -0.00440 \\
  0.0675 &  -0.0102 \\
  0.0899 &  -0.0219 \\
   0.1167  &   -0.0440 \\
   0.1484  &    -0.0834 \\
   0.160   &    -0.102 \\
   0.166    &   -0.113 \\
   0.185   &    -0.151 \\
   0.228     &  -0.262 \\
\hline
\hline
\end{tabular}
\end{table*} 

In Fig.~\ref{Kaiser1}, the blue curve shows the contribution to the energy per nucleon as a function of density in Born approximation from the contact terms at N$^4$LO, Eq.~(\ref{EoA_kaiser}). The red is the contribution in Born approximation from the contact term at N$^2$LO with $c_E$ = -1.27, as obtained from the 3N continuum, while the green curve represents the same quantity as the red one, but with $c_E$ = 0.13~\cite{WGS22}. The incompatibility between the contact strengths obtained through the 3N continuum and the energy of SNM is clear. We anticipate a similar scenario with the binding energy of nuclei.

From naive dimensional considerations, one might expect the contribution to the energy per nucleon from the contact in-medium potential at N$^4$LO to be suppressed, roughly, by a factor of $(k_F/ \Lambda)^2$ with respect to the leading contact contribution. Applying this prescription, we obtain the values shown in Table~\ref{supr}, which may be seen as some sort of guideline or loose constraint from the nuclear matter sector.
\begin{figure*}[!t] 
\centering
\hspace*{-0.5cm}
\includegraphics[width=5.5cm]{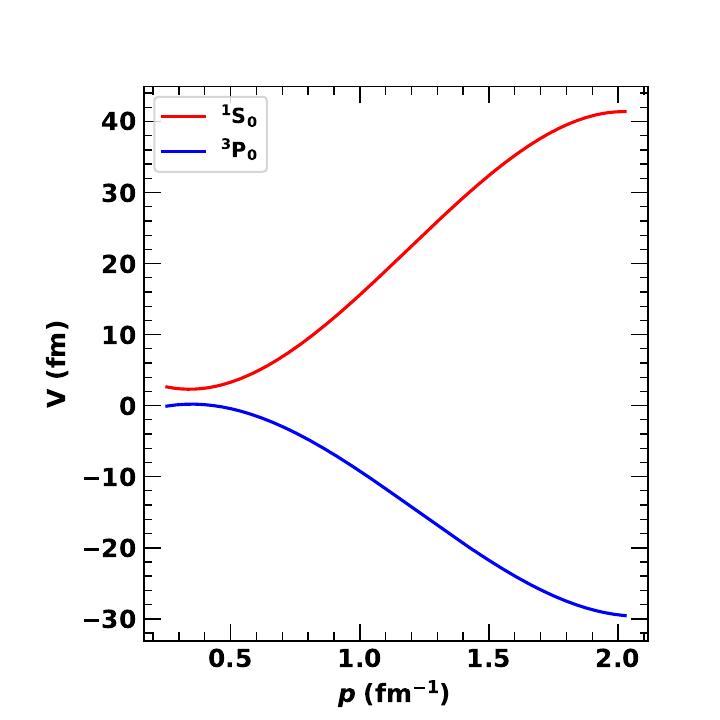}\hspace{0.01in}  
\includegraphics[width=5.5cm]{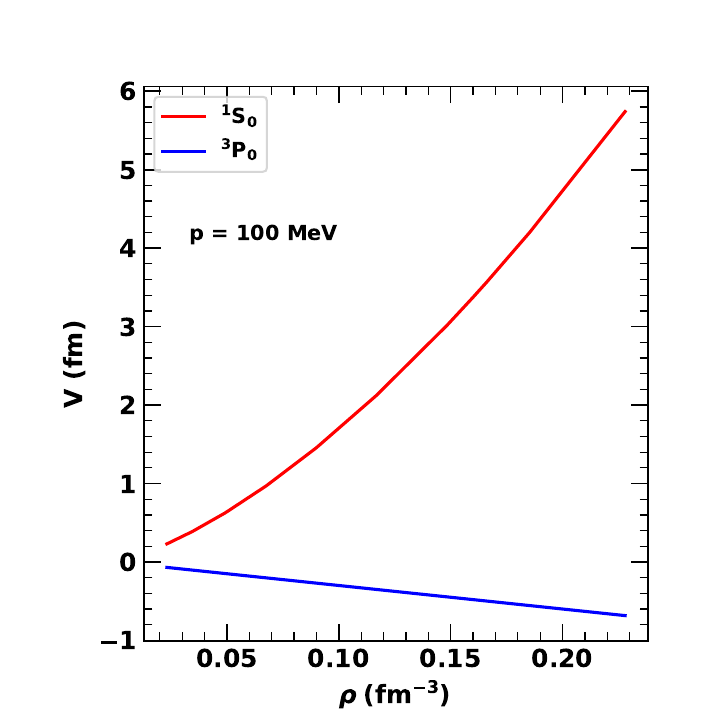}\hspace{0.01in}  
\includegraphics[width=5.5cm]{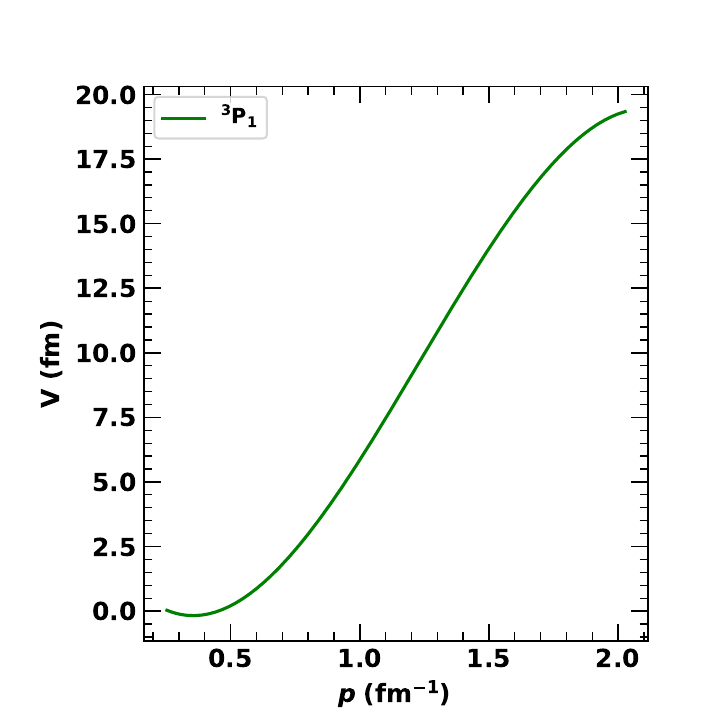}\hspace{0.01in}  
\includegraphics[width=5.5cm]{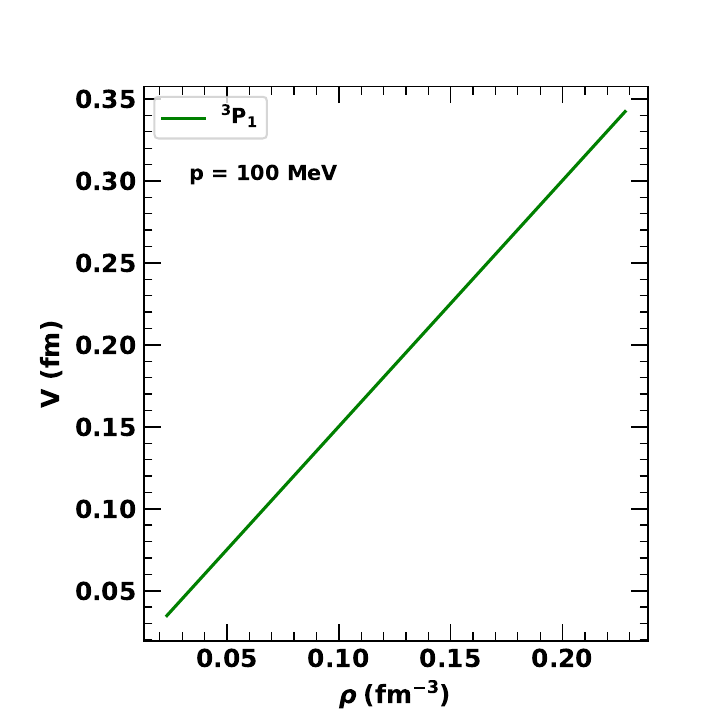}\hspace{0.01in} 
\vspace*{-0.1cm}
 \caption{First two frames from the left: Matrix elements of the potential in Eq.~(\ref{vmedn}) for J=0 at normal density as a function of the momentum, and as a function of density for fixed momentum. Next two frames: Same legend, for the allowed state with J=1.
}
\label{nm1}
\end{figure*}   
\section{Neutron matter}
\label{NM}

The contributions from the $c_D$ and $c_E$ couplings vanish in the pure neutron system, which makes understanding of the contact at N$^4$LO especially important.

The subleading three-neutron (3n) contact potential can be written as~\cite{Kai20_nn}:
\begin{equation}
\label{vmedn}
\begin{split}
\frac{V_{med}}{\rho} & = 2(E_1 + E_2) \big (\frac{6}{5} k_{f,n}^2 +2 p^2 -q^2 \big ) + 2(E_3+E_4) ( \vec{\sigma}_1 \cdot \vec{\sigma }_2 + 3)  \big (\frac{3}{5} k_{f,n}^2 + p^2 \big)  \\
& + 3(E_5+E_6) \big [ \vec{\sigma}_1 \cdot \vec{p} \vec{\sigma}_2 \cdot \vec{p} + \vec{\sigma}_1 \cdot \vec{p'} \vec{\sigma}_2 \cdot \vec{p'} -\frac{2p^2}{3} \vec{\sigma}_1 \cdot \vec{\sigma}_2   \big ]  \\
& + (E_7 + E_8)i( \vec{\sigma}_1 + \vec{\sigma }_2)\cdot (\vec{q} \times \vec{p}) \\
&  + (E_9 + E_{10})\big [q^2 -2p^2 - \frac{6}{5} k_{f,n}^2 -i  ( \vec{\sigma}_1 + \vec{\sigma }_2)\cdot (\vec{q} \times \vec{p})  \big ]  \\
& + (E_{11} + E_{12} + E_{13})\big [q^2 -2p^2 - \frac{6}{5} k_{f,n}^2 +i ( \vec{\sigma}_1 + \vec{\sigma }_2)\cdot (\vec{q} \times \vec{p})  \big ]  \; ,
\end{split}
\end{equation}
where $k_{f,n}$ is the fermi momentum in NM at density $ \rho = k_{f,n}^3/3 \pi^2$
from which one can derive 
 the contribution to the energy per neutron at first order~\cite{Kaiser21}:
\begin{equation}
\label{EoN_kaiser}
\frac{E}{N} = \frac{k_{f,n}^8}{30 \pi ^4} (2E_1+2E_2 -E_9 -E_{10}-E_{11}-E_{12}-E_{13}) \; .
\end{equation}

\begin{table*}
\caption{First order contribution to the energy per neutron in neutron matter at N$^4$LO from Eq.~(\ref{EoN_kaiser}). }
\label{EN}
\centering
\begin{tabular}{|l|r|}
\hline
\hline
$\rho$ ($fm^{-3}$)  &  E/A (MeV) \\
\hline
0.0232  &  0.470 \\
  0.0346 &  1.365 \\
  0.0492 &  3.490 \\
  0.0675 &  8.112 \\
  0.0899 &  17.417 \\
   0.1167  &   34.926 \\
   0.1484  &    66.290 \\
   0.160   &    81.024 \\
   0.166    &   89.382\\
   0.185   &    119.330 \\
\hline
\hline
\end{tabular}
\end{table*} 
 At this time, the same conclusions as for SNM apply with regard to the excessively large size of these contributions.

\section{Finite nuclei}

As nuclear matter is the result of an extrapolation from finite nuclei, the saturation properties of SNM, that is, the minimum of the EoS at the appropriate equilibrium density, should be naturally related
to the energy and density distributions of nucleons in nuclei. Typically, the extrapolation to SNM
has been done with phenomenological density-dependent forces, such as Skyrme or Gogny forces~\cite{Brown13, Mon18, DG80}. Applying such
forces, it has been observed that a good phenomenological description of nuclei extrapolates to a saturation density
in nuclear matter of about 0.16 fm$^{-3}$ and energy per particle of approximately -16 MeV.

To establish a direct link between the EoS and nuclear binding energy, we calculate the latter with an intuitive picture, established in nuclear physics since decades, that describes a nucleus in terms of a mass formula,
whose extrapolation to an infinite electrically neutral system defines nuclear matter. Although simple, this model
is not fundamentally wrong, especially for bulk properties such as energies and r.m.s. radii, namely averaged
values rather than quantum structures. 

The main contribution to the mass formula comes from the energy per particle in asymmetric matter, for which we use the
expansion quadratic in $\alpha$, where $\alpha = \frac{\rho_n - \rho_p}{\rho_n + \rho_p}$ is the isospin asymmetry:
\begin{equation}
\label{alpha}
e(\rho, \alpha) = \approx e_0(\rho) + e_{sym}(\rho) \alpha^2 \; .
\end{equation} 
In  Eq.~(\ref{alpha}), $\rho$ is the total nucleon density and  $e_{sym}(\rho)$ is the symmetry energy.

\begin{figure*}[!t] 
\centering
\hspace*{-0.5cm}
\includegraphics[width=7.0cm]{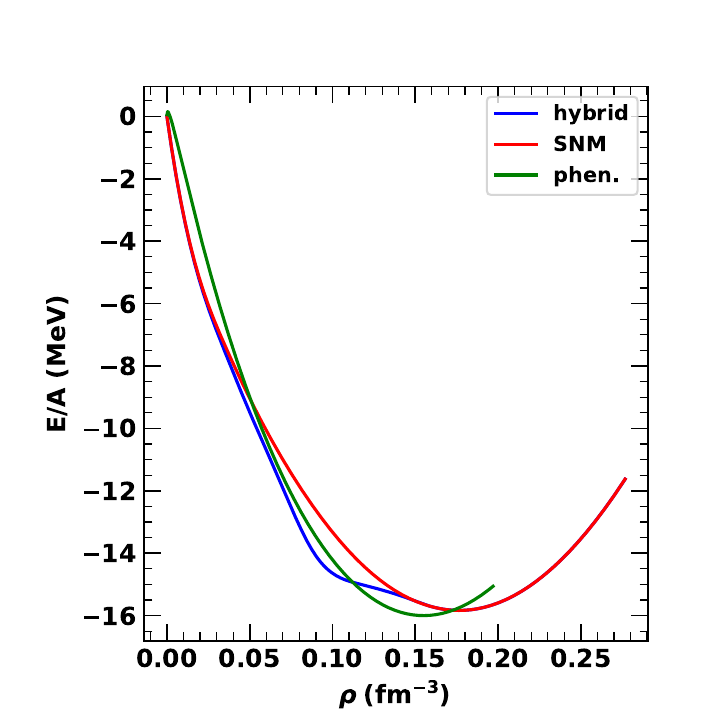}\hspace{0.01in}  
\vspace*{-0.1cm}
 \caption{Red: Our standard EoS at N$^2$LO with the N2LO450 potential~\cite{EMN17}, $c_D$ = 2.75, and $c_E$ = 0.13; Blue: EoS at N$^2$LO with the N2LO450 potential, $c_D$ = 6.0, and $c_E$ = 0.50 (see text for explanation); Green: A  phenomenological parametrization of the EoS~\cite{Oya+2010}.
}
\label{hybrid}
\end{figure*}

The following is an exploratory test to probe the sensitivity of the binding energy per nucleon (BE/A) to the behavior of the EoS in specific density regions.
From the $c_D$, $c_E$ trajectory that fits the triton~\cite{DHS19}, we have selected the largest pair of positive values, namely $c_D$ = 6.0 and $c_E$ = 0.50, and used it to calculate the EoS up to $\rho \approx $ 0.12 fm$^{-3}$, blue curve in Fig.~\ref{hybrid}, at which point we rejoin the EoS obtained with $c_D$ = 2.75 and $c_E$ = 0.13, red curve in Fig.~\ref{hybrid}. The outcome is additional attraction at densities between 0.07 and 0.12 fm$^{-3}$. Clearly, continuing the blue curve to higher densities would prevent a reasonable saturating behavior. The green curve is a phenomenological EoS~\cite{Oya+2010}, as an additional element of comparison.

Looking at Fig.~\ref{hybrid} together with Table~\ref{EB}, we observe that: 
\begin{enumerate}
\item The red curve saturates the EoS with energy of -16 MeV  at a larger density than the phenomenological parametrization (green). It underbinds the nuclei in Table~\ref{EB} due to insufficient attraction at the appropriate density.
\item The green curve demonstrates what could be an ideal saturation point, at 0.155 fm$^{-3}$  and E/A = -16 MeV. The shift of the minimum to lower densities as compared to the red EoS amounts to extra attraction in the region where it's needed to enhance the binding of nuclei. 
\item With a bold change in the short-range couplings (while still binding the triton), the blue EoS simulates the attraction needed to get close to the experimental BE/A values, especially Calcium.
\end{enumerate}

\begin{table*}
\caption{BE/A in MeV for the given nuclei evaluated with the red, blue, and green EoS from Fig.~\ref{hybrid}.
}
\label{EB}
\centering
\begin{tabular*}{\textwidth}{@{\extracolsep{\fill}}|ccccc|}
\hline
\hline
nucleus & BE/A (Experimental) & BE/A (red curve) & BE/A (blue curve) & BE/A (green curve) \\
\hline
$^{16}$O  & 7.98 & 7.06 & 7.88 & 7.41  \\
$^{40}$Ca  & 8.55 & 7.91 & 8.56 & 8.33  \\
$^{208}$Pb  & 7.87  & 7.22 & 7.72 & 8.17  \\
\hline
\hline
\end{tabular*}
\end{table*} 

\section{Some recent developments}

In this section, we address recent findings reported in Refs.~\cite{Sola+26, FGC26}. There, a spectroscopic basis is introduced specifically to represent
 the subleading contact 3NF. To define the spectroscopic
basis for the terms comprising the most general structure of the subleading contact 3NF in operator basis, Sola$'$ {\it et al.} performed
an analytic partial-wave decomposition of the complete antisymmetrized 3NF and determined that only few linear combinations
of the original $E_i$ contribute to a specific $J^P(T)$-channel. They then defined a new set of LECs,
named $S_i$ (or $s_i$ if in the dimensionless form), as linear combinations of the $E_i$. This new basis has the natural advantage of providing a more transparent connection between the 3N observables and the contact operator terms that contribute to specific partial waves~\cite{Sola+26}. These findings are reported to be consistent with those from Ref.~\cite{FGC26}.

We performed the inverse linear transformation provided in Ref.~\cite{Sola+26} and extracted the $e_i$ shown in Table~\ref{lecs}. The LECs of the leading contact 3NF were also refitted after inclusion of the subleading contact~\cite{Sola+26} and are included in Table~\ref{lecs} as well. To extract uncertainties for the $E_i$, while solving the system in Eq.~(C.1) of Ref.~\cite{Sola+26} we allowed the $S_i$ to vary independently within their respective uncertainties and obtained the values in Table~\ref{lecs}. Extending our
discussion of Fig.~\ref{Kaiser1}, we notice the positive value and the smaller magnitude for $c_E$, to be compared with $c_E$=-1.27 $\pm$ 0.06. 

\begin{figure*}[!t] 
\centering
\hspace*{-0.5cm}
\includegraphics[width=7.0cm]{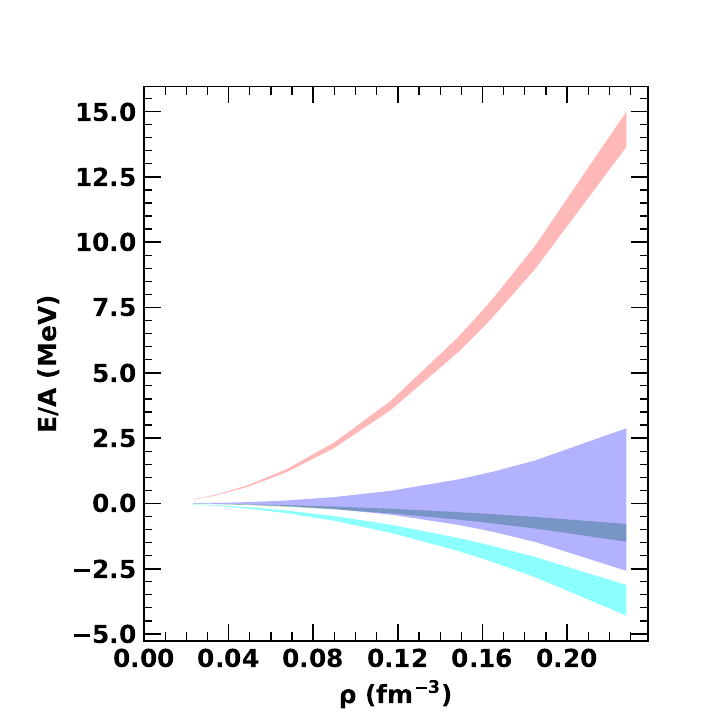}\hspace{0.01in}  
\vspace*{-0.5cm}
 \caption{Purple band: Born approximation contribution to the energy per particle in SNM from the subleading 3NF contact potential, obtained with the $E_i$ LECs based on Ref.~\cite{Sola+26}. 
Green (within the purple band): Leading contact (N$^2$LO) with $c_E$=(0.07, 0.130)~\cite{DHS19}; Cyan: Leading contact with $c_E$=0.329 $\pm$ 0.052; Pink: Leading contact with $c_E$=-1.27 $\pm$ 0.06. }
\label{sola}
\end{figure*}

We then propagated those errors into the leading and subleading contact contributions to the energy per particle, see Fig.~\ref{sola}.
 The green, cyan, and pink bands represent the first-order contributions to the energy per nucleon in SNM from the leading contact 3NF, using $c_E$ = 0.130~\cite{DHS19}, 0.329~\cite{Sola+26}, and -1.27~\cite{WGS22}, respectively. With the small and positive values of $c_E$, the contribution to the energy per particle in Born approximation is attractive and has a magnitude between about 1.5 and 2 MeV at saturation. The blue curve shows the contribution from the subleading contact potential with the $E_i$ strengths from Table V, right column, which we obtained based on Ref.~\cite{Sola+26}.

\begin{table*}
\caption{The first two rows display the LECs of the leading short-range 3NF. Rows 3 through 15: couplings of the contact contribution to the 3NF at N$^4$LO. All couplings are dimensionless. }
\label{lecs}
\centering
\begin{tabular}{|l|r|r|}
\hline
\hline
LEC & Ref.~\cite{WGS22} &  From or based on \\
     &                                   & Ref.~\cite{Sola+26} \\
\hline
$c_D$  &  -1.49 $\pm$ 0.06 & -2.510 $\pm$ 0.385 \\ 
$c_E$   & -1.27 $\pm$ 0.06 & 0.329 $\pm$ 0.052 \\
\hline
 $e_1$  &  6.40 $\pm$ 0.33 & (-1.240, -0.0687 \\
  $e_2$  &  7.80 $\pm$ 0.36 & (-3.147, -1.0118) \\
$e_3$  &  6.97 $\pm$ 0.34 & (1.30, 3.795) \\
$e_4$  &  -2.06 $\pm$ 0.13 & (-0.315, 0.583) \\
$e_5$  &  -0.36 $\pm$ 0.05 & (1.993, 4.305) \\
  $e_6$  &  0.52 $\pm$ 0.03 & (0.331, 1.364) \\
$e_7$  &  -7.40 $\pm$ 0.14 & (-4.759, 0.117) \\
$e_8$  &  -2.61 $\pm$ 0.05 & (-1.920, -0.177) \\
$e_9$  &  -4.59 $\pm$ 0.22 & (-1.201, 2.503) \\
 $e_{10}$  &  -0.98 $\pm$ 0.05 & (-0.921, 0.891) \\
$e_{11}$  &  -4.59 $\pm$ 0.22 & (-2.078, 1.181) \\
$e_{12}$  &  -0.98 $\pm$ 0.05 & (-2.127, -0.571) \\
$e_{13}$  &  -1.14 $\pm$ 0.05 & (-6.613, -2.184) \\
\hline
\hline
\end{tabular}
\end{table*} 

A positive and natural value for $c_E$ appears to be favored by the 3N continuum, the 3N bound state, and (potentially) nuclear matter saturation. With regard to the subleading contribution, we notice that the $E_i$ values in the third column of Table~\ref{lecs} display, overall, a more natural magnitude than those on the left. A striking feature we observe is the strong suppression of the N$^4$LO contribution to the energy per nucleon in SNM, as can be seen comparing the blue curves in Fig.~\ref{Kaiser1} and in Fig.~\ref{sola}.
The magnitude of this contribution remains small, but attraction or repulsion cannot be determined within the current uncertainty.  We emphasize that these are preliminary observations. Additional Nd scattering measurements~\cite{Saito+26} have the potential to provide most valuable information.

An important comment is in place. We have set $s_2$ and $s_4$ to zero, because they were found redundant for the purpose of fitting elastic Nd scattering observables. In fact, varying these couplings within a reasonably natural range was found to change the $\chi^2$/datum by less than 1\% as compared to its value for $s_1$=$s_2$=0~\cite{Sola+26}. The $s_{12}$ and $s_{13}$ LECs were discarded from the fit because they contribute only to the T=3/2 channel and, thus, cannot be determined {\it via} elastic Nd scattering. We have set these couplings equal to zero as well, with the understanding that their values cannot be constrained at this stage.

We found the contribution to neutron matter from the subleading contact 3NF obtained from the spectroscopic strengths to be vanishingly small at first order, but this observation is under current investigation.

\section{Takeaways}

Our main takeaways and open questions can be summarized as follows. The values of $c_D$, $c_E$ and $E_i$ depend dramatically on the system from which they are extracted, because different systems probe different densities.
Ideally, one wants to reconcile the 3N bound state, nuclear matter and medium-mass nuclei, and the 3N continuum, or, at the very least, understand the origin of these irregularities. As the distribution of momenta depends on the system under consideration, a possible candidate is the amount of off-shell component brought in by these couplings.

Of course, the irregularities associated with the $c_D, c_E$ LECs were known prior to the fitting of the $E_i$ couplings {\it via} 3N scattering, driving the anticipation that the 3NF contact at N$^4$LO might provide some answers.

The ideal combination of 2NF and 3NF must provide additional attraction at densities more typical of the average density in nuclei (rather than their central density). Because 3NF have minor impact at low densities, it would appear that the 2NF should be considerably softer at low-medium density.

Recent fittings of 3N observables using a spectroscopic instead of an operator basis~\cite{FGC26, Sola+26}, provide values of
the $E_i$ LECs that are different from those extracted in Ref.~\cite{WGS22}. These new couplings yield contributions to the first-order energy per nucleon in SNM from the subleading 3NF potential that differ by orders of magnitude from those shown in Fig.~\ref{Kaiser1}, blue curve. We also note that the blue band in Fig.~\ref{sola} is not inconsistent with the rough estimates in Table~\ref{supr} based on naive dimensional arguments.

\section*{Acknowledgments}
This work was supported by 
the U.S. Department of Energy, Office of Science, Office of Basic Energy Sciences, under Award Number DE-FG02-03ER41270.

\bibliography{bibRM}

\end{document}